# Predicting Hidden Links and Missing Nodes in Scale-Free Networks with Artificial Neural Networks

Rakib Hassan Pran [1]
Web of Science ResearcherID: JDC-3499-2023

[1] M.Sc. in Applied Statistics with Network Analysis,
International Laboratory for Applied Network Research [2 3 4]

[2] was established under the leadership/scientific supervision of Stanley Wasserman, the Rudy Professor Emeritus of Psychological and Brain Sciences, Statistics, and Sociology, Indiana University Bloomington, US
[3] was conducted in scientific collaboration with professors from faculties, University of Ljubljana, SI
[4] was organized by National Research University Higher School of Economics, RU

## Abstract

**There are many networks in real life which exist as form of Scale-free networks such as World Wide Web, protein-protein inter action network, semantic networks, airline networks, interbank payment networks, etc. If we want to analyze these networks, it is really necessary to understand the properties of scale-free networks. By using the properties of scale free networks, we can identify any type of anomalies in those networks. In this research, we proposed a methodology in a form of an algorithm to predict hidden links and missing nodes in scale-free networks where we combined a generator of random networks as a source of train data, on one hand, with artificial neural networks for supervised classification, on the other, we aimed at training the neural networks to discriminate between different subtypes of scale-free networks and predicted the missing nodes and hidden links among (present and missing) nodes in a given scale-free network. We chose Bela Bollobas's [2] directed scale free random graph generation algorithm as a generator of random networks to generate a large set of scale-free network's data.**

## 1. Introduction

Scale-free networks follows the power law degree distribution [3]. In scale free network, the fraction $P(k)$ of nodes in the network which has $k$ connections to other nodes follows the distribution

$$P(k) \sim k^{-\Upsilon}$$

Or,

$$\frac{P(k)}{k^{-\Upsilon}} = 1$$

Where, $\Upsilon$ is a real valued parameter, $\Upsilon > 0$; generally the value of $\Upsilon$ is in the range $2<\Upsilon<3$.

$$P(k) = \frac{number\ of\ nodes\ having\ k\ connection}{total\ number\ of\ nodes}$$

Directed scale-free networks are scale free networks which have directed edges and the in-degree distribution and out-degree distribution of directed scale-free networks both follow power law.

Scale-free networks are ubiquitous, as they take many forms in real life. For example, many internet networks follow the power law degree distribution such as web graph of World Wide Web [4], connection of IP addresses in internet [5], etc. Besides that, software dependency graphs [6], P2P client server model graphs [7], worm propagation [8], semantic network in Natural Language Processing [9] and many other networks in computer science follows the properties of scale-free networks. In biology, E. coli metabolism [10], Protein-Protein interaction [11] are also scale-free networks. Besides that, some financial networks such as interbank payment networks [12] [13], airline networks [14], the collaboration of actors in movies [15], the co-authorship of research papers [16] also follows power law degree distribution as scale-free networks.

But the questionable part for scale free network is if they exist in real life as many forms then the first question is- what kind of attributes or properties they are possessing to form same kind of network and the second question is- what kind of differences can be exist in scale free network though they all follows power law degree distribution.

To understand these two questions, we analyzed the parameters of Bela Bollobas's [2] directed scale free random graph generation algorithm which we chose as a generator of random networks (described in section 2). By analyzing these parameters, we figured out, scale-free networks follow the range of each parameters. In this case, to figure out the range of each parameter, we followed Pran's "Correlation analysis among social network measures for directed scale free network" [17]. After getting the range for each parameters, we followed Guzman's "An Analytical Comparison of Social Network Measures" [18] to choose the sample from the range of each parameter. Then we used these samples in random graph generator algorithm [2] to create different sub-types of scale-free networks and labeled those networks as per the sub-types of scale-free networks to train our first artificial neural network. After proper training of our first Artificial Neural Network (by generating a large number of scale-free networks and labeled them as per the sub-types of scale-free networks), we used this artificial neural network to predict the parameters of given scale-free network (on which, we want to predict the hidden links and missing nodes). Then, we used this predicted value of parameters in random graph generation algorithm [2] to generate a large number of random scale-free networks which all are the same sub-type of the given scale-free network. We used these large number of similar sub-type of given scale-free network (labeled as valid) and other sub-types of scale-free networks (labeled as invalid) to train our second Artificial Neural Network. Now, If the given scale-free network is not large, we will create networks as test data by putting all possible combination of 1's and 0's in positions of 0's (but not position of 1's) of given scale-free network's adjacent matrix, otherwise, If the given scale-free network is large, we will take random sample from the combination of 1's and 0's in positions of 0's in adjacent matrix. We considered networks as predicted networks with hidden links and missing nodes for given scale-free network if they got prediction accuracy greater than 0.80 for valid network group.

## 2. Generating random scale-free networks:

In this Model,

α , β, ϒ, δin and δout are 5 parameters where

$\alpha + \beta + \gamma = 1 \ldots\ldots\ldots\ldots\ldots eq(1)$

G(n) is a directed random network with n edges and N(n) nodes.

Let us denote, the set of nodes of G (n) with Vn ; |Vn| = N (n) and

The set of edges of G (n) with En, a suset of { Vn x Vn}

We will denote the In-degree of v with $D_{in}$(v)

& out-degree of v with $D_{out}$(v).

With the probability α, append a new node to G (n-1), v$\notin V_{n-1}$ and create directed edge v→w $\in V_{n-1}$ with probability of

$$\frac{D_{in}(w)+\delta_{in}}{n-1+\delta_{in}N(n-1)}$$

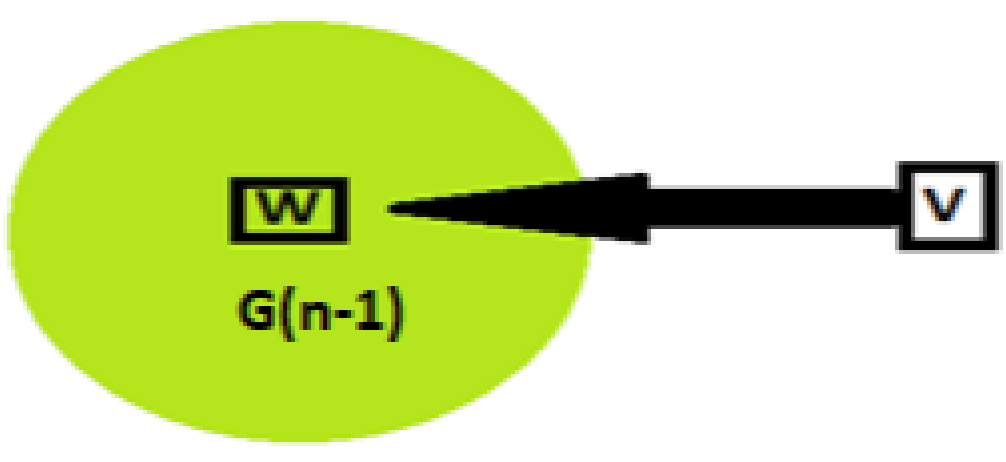

With probability γ, append to G (n-1), a new node, v$\notin V_{n-1}$ and create directed edge w$\in V_{n-1}$ → v$\notin V_{n-1}$ with probability of

$\frac{D_{out}(w)+\delta_{out}}{n-1+\delta_{out}N(n-1)}$

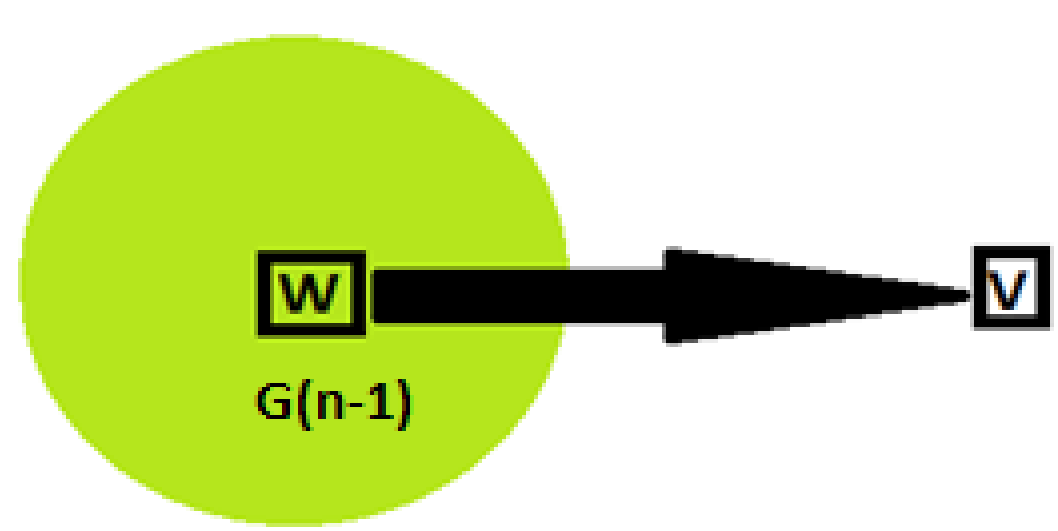

With probability β, creating new directed edge between existing nodes.

Here, newly created edge v $\in V_{n-1}$ $\rightarrow$ w $\in V_{n-1}$ with probability of $\left(\frac{D_{in}(w)+\delta_{in}}{n-1+\delta_{in}N(n-1)}\right)\left(\frac{D_{out}(w)+\delta_{out}}{n-1+\delta_{out}N(n-1)}\right)$

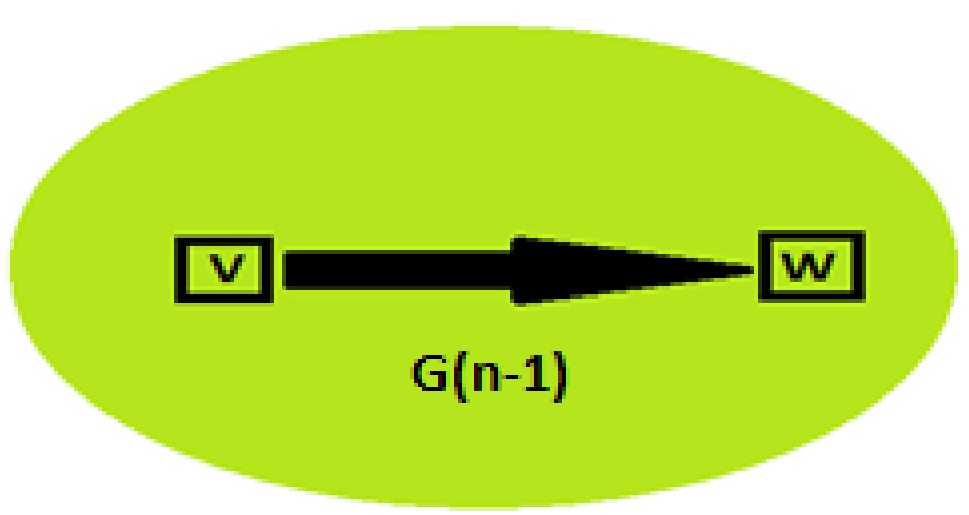

The proportion of nodes with in-degree is

$$p_i = \frac{N_i(n)}{N(n)}$$

Where, Ni (n) is the number of nodes with in-degree i

The proportion of nodes with in-degree j is

$$q_j = \frac{N_j(n)}{N(n)}$$

Where,

Nj (n) is the number of nodes with out-degree j

The $p_i$, which has power law tail, so we got,

$$p_i \sim C_{IN} i^{-X_{in}}$$

Where,

$C_{IN}$ Has positive constant and $i \to \infty$ and $\alpha\delta_{in} + \gamma > 0$ and

$X_{in} = \frac{\alpha+\beta+1+\delta_{in}(\alpha+\gamma)}{\alpha+\beta}, \delta_{in} \geq 0$...........$eq(2)$

And

$\delta_{in} = \frac{X_{in}(\alpha+\beta)-\alpha-\beta-1}{\alpha+\gamma}, \delta_{in} \geq 0$...........$eq(3)$

The $q_i$ has power law tail so we got,

$$q_i \sim C_{out} i^{-X_{out}}$$

Where,

$C_{out}$ Has positive constant and $i \to \infty$ and $\gamma\delta_{in} + \alpha > 0$ and

$X_{out} = \frac{\gamma+\beta+1+\delta_{out}(\alpha+\gamma)}{\gamma+\beta},$

$\delta_{out} \geq 0$...........$eq(4)$

And

$\delta_{out} = \frac{X_{out}(\gamma+\beta)-\gamma-\beta-1}{\alpha+\gamma},$

$\delta_{out} \geq 0$...........$eq(5)$

So, we can see from this random graph generation model, there are 5 parameters (α, β, ϒ, δin and δout) which are used to generate different types of Scale free network. These 5 parameters can vary to different types of scale free network though their in-degree distribution, $X_{in}$ and out-degree distribution,$X_{out}$ range are between (2, 3).

So, we got, generally, α, β, ϒ, δin and δout can be different for different types of scale free network but in-degree distribution, $X_{in}$ and out-degree distribution,$X_{out}$ range will always be within 2 to 3 for scale free network.

Now, we are going to use this random graph generation algorithm to create our own algorithm for estimating hidden links and missing nodes in directed scale free network by using artificial neural network system which is described in methodology section. At the end of Methodology section, we stated our pseudo code of our algorithm.

## 3. Artificial neural networks for scale-free networks

To estimate missing links and hidden nodes in scale free network, we implemented an algorithm where we used a sequential model algorithm of artificial neural network and Bela Bollobas's directed scale free random graph generation algorithm [2] by optimizing its 5 parameters. After describing the whole algorithm, at the end of the methodology, the pseudo code of this algorithm will be stated.

First, we need to convert the considered directed scale free network (on which we want to estimate the hidden links and missing nodes) into adjacent matrix.

Let, assume the selected directed scale free network (on which we want to estimate the hidden links and missing nodes) is $G$ and the adjacent matrix of that directed scale free network is $G_N$ where N is the number of nodes in network and size of the $G_N$ is $N\times N$ matrix . If we want to estimate $M$ number of missing nodes in $G_N$ then we will add $M$ unconnected nodes in $G_N$. So, let assume, after adding $M$ unconnected nodes in $G_N$, the newly created network adjacent matrix is $G_{new}$ where, $N + M$ is the number of nodes in $G_{new}$ and the size of $G_{new}$ is $(N + M)\times(N + M)$ matrix.

Now, we need to generate a large set of $(N + M)\times(N + M)$ matrix sized directed scale free networks to train data and label them as per their types. After that, we will use this labeled training data to train our artificial neural network model.

To generate this large set of $(N + M)\times(N + M)$ matrix sized directed scale free networks, we used Bela Bollobas's random graph generation algorithm for directed scale free network. In this algorithm, there are 5 parameters which are α, β, ϒ, δin and δout. From previous research "Correlation analysis among social network measures for directed scale free network" [17], we have seen that, in this algorithm if we consider directed scale free network where $X_{in}$, $X_{out}$ ranges are between 2.1 to 3.0 then the range of 5 parameters will be α=(0.1,0.5), β =(0.1,0.8), ϒ =(0.1,0.5), δin=(0.0,4.0) and δout=(0.0,4.0).

Now, by using the combination of α, β, ϒ, δin and δout within these ranges, we can generate different types of directed scale free network.

Let, take the set of α, β, ϒ, δin and δout where

α = {0.1,0.2,0.3,0.4,0.5}

β = {0.1,0.2,0.3,0.4,0.5,0.6,0.7,0.8}

ϒ = {0.1,0.2,0.3,0.4,0.5}

δin = {0.0,0.2,0.4,0.6,…….,3.6, 3.8, 4.0}

δout = {0.0,0.2,0.4,0.6,…….,3.6, 3.8, 4.0}

But, instead of δin and δout, we are going to take with respect to Xin and Xout because of our research procedure. It's possible to take Xin and Xout instead of δin and δout because the relation between δin and Xin exist as like as δout and Xout which are considered as the properties of Bela Bollobas's directed scale free network graph generation algorithm.

The equations for the properties of Bela Bollobas's directed scale free network graph generation algorithm are $eq(1)$, $eq(2)$ $or$ $eq(3)$ $and$ $eq(4)$ or $eq(5)$

We can see, we may use eq(2) instead of eq(3) and eq(4) instead of eq(5)

So, Instead the range of δin=(0.0,4.0) and δout=(0.0,4.0), we are going to use,

The range of Xin = (2.1, 3.0) and Xout = (2.1, 3.0).

So, newly constructed set are

α = {0.1,0.2,0.3,0.4,0.5}

β = {0.1,0.2,0.3,0.4,0.5,0.6,0.7,0.8}

ϒ = {0.1,0.2,0.3,0.4,0.5}

$X_{in}$ = {2.1, 2.2, 2.3, ….., 2.9, 3.0}

$X_{out}$ = {2.1, 2.2, 2.3, ….., 2.9, 3.0}

## 3.1 ANN for discriminating between small-scale subtypes

Now, we are going to use the combination of α, β, ϒ, $X_{in}$ and $X_{out}$ to generate large set of $(N + M)\times(N + M)$ matrix sized directed scale free network but we labeled each network as per the combination of $X_{in}$ and $X_{out}$.

There are 10 elements in $X_{in}$ and 10 elements in $X_{out}$.So, the number of group in labeled data will be $10C_1 \times 10C_1$= 100. So, the set of Groups is $\{Group1, Group2, Group3, ...., Group100\}$

For each group, we generated $N + M$ number of directed scale free network where each network has $N + M$ number of nodes because, to predict "estimating hidden links and missing nodes in large network", the network with large number of nodes needs large number of trained network data. So, in training data, the number of network will increase as per the number of network nodes.

So, the total number of directed scale free network in training data is $100(N + M)$ which are $\{G1, G2, G3, ...., G(100(N + M))\}$

As Artificial Neural Network, we are using sequential model let, assume $ANN1$ where we implemented one input layer, two hidden layers and one output layer.

We need to take complete $G_{new}$ graph as Input layer at the time of doing test. So, Input layer size will be $(N + M)\times(N + M)$ matrix. So, the total number of nodes in input layer is $(N + M)\times(N + M)$.

The number of nodes of first hidden layer is $\frac{(N+M)\times(N+M)+100}{2}$ and the number of second hidden layer is $\frac{\frac{(N+M)\times(N+M)+100}{2}+100}{2}$

As activation function of first and second hidden layers, we are using Rectifier linear function, $f(x) = x^{+} = (0, x)$ , $where,$ $x$ is the input to a neuron.

The number of nodes of output layer is $100$, each output neuron correspond to one of the groups from $\{Group1, Group2, Group3, ...., Group100\}$

And as activation function of output layer, we are using $Softmax$ function,

$$Softmax\left(z_i\right) = \frac{e^{z_j}}{\sum_{k=1}^{K} e^{z_k}} \quad for\ z = 1, 2, ..., 100$$

Which is a standard activation function in the output layer of the neural networks for classification. It takes as input a vector *z* of *K* real numbers and normalizes it into a probability distribution consisting of *K* probabilities which is proportional to the exponentials of the input numbers. It is often used in neural networks to map the non-normalized output of a network for a probability distribution over predicted output classes.

So, we trained this artificial neural network model $ANN1$ by training data $\{G1, G2, G3, ...., G(100(N + M))\}$ where the number of training data is $100(N + M)$ and each training data is nothing but labeled directed scale free network.

Now, we use $G_{new}$ as test data to predict the Group it belongs. In this case, we took the Group which has the highest prediction accuracy for $G_{new}$. We are choosing the highest prediction accuracy to use highly accurate values of parameters which will be used in random graph generation algorithm to generate the closest accurate scale-free networks data for $G_{new}$ to train next artificial neural network. Let us, assume that the predicted Group for the highest prediction accuracy for $G_{new}$ is

$Group_{predicted}$ $where\ \ Group_{predicted} \in \{Group1, Group2, Group3, ...., Group100\}$,

Which has the highest accuracy for $G_{new}$ among all these 100 groups and

Let, assume the $X_{in}$ and $X_{out}$ of $Group_{predicted}$ is sequentially $X_{in_{predicted}}$ and $X_{out_{predicted}}$

## 3.2 ANN for predicting missing nodes and hidden links

By predicting, $Group_{predicted}$, now, we know about the $G_{new}$'s $X_{in} = X_{in_{predicted}}$ and $G_{new}$'s $X_{out} = X_{out_{predicted}}$

Now, we are going to create another Sequential Artificial Neural Network Model $ANN2$ where one input layer, one hidden layer and one output layer.

The total number of nodes in input layer is $(N + M)\times(N + M)$

The total number of nodes in hidden layer is $\frac{(N+M)\times(N+M)+2}{2}$

As activation function of first hidden layers, we are using Rectifier linear function, $f(x) = x^{+} = (0, x)$ , $where,$ $x$ is the input to a neuron.

And the total number of output layer is 2 which are $Group_{valid\ network}$ and $Group_{invalid\ network}$

To train this ANN2, we need to create different training data where directed scale free network's $X_{in} = X_{in_{predicted}}$ and $X_{out} = X_{out_{predicted}}$.

So,

α = {0.1,0.2,0.3,0.4,0.5}

β = {0.1,0.2,0.3,0.4,0.5,0.6,0.7,0.8}

ϒ = {0.1,0.2,0.3,0.4,0.5}

$X_{in} = \{X_{in_{predicted}}\} U\{ \{2.1,\ 2.2,\ 2.3,\ .....,\ 2.9,\ 3.0\} - \{X_{in_{predicted}}\}\}$

$X_{out} = \{X_{out_{predicted}}\} U\{ \{2.1,\ 2.2,\ 2.3,\ .....,\ 2.9,\ 3.0\} - \{X_{out_{predicted}}\}\}$

We used the combination of α, β, ϒ, $X_{in}$ and $X_{out}$ to generate a large set of $(N + M)\times(N + M)$ matrix sized directed scale-free networks training data.

To label these network, we label network as $Group_{valid\ network}$ if $X_{in} = X_{in_{predicted}}$ and $X_{out} = X_{out_{predicted}}$ otherwise network will be labeled as $Group_{invalid\ network}$

To predict hidden links and missing nodes in $G_{new}$, we need to test all possible links for $G_{new}$ but among all these possible links only few will satisfy the property of $X_{in} = X_{in_{predicted}}$ and $X_{out} = X_{out_{predicted}}$; and those networks will be in $Group_{valid\ network}$. For checking this, we will put α= {0.1,0.2,0.3,0.4,0.5}, β={0.1,0.2,0.3,0.4,0.5,0.6,0.7,0.8}, ϒ={0.1,0.2,0.3,0.4,0.5}, $X_{in} = X_{in_{predicted}}$ and $X_{out} = X_{out_{predicted}}$ in $eq(3)\ and\ eq(5)$ where, δin and δout should be in range of (0.0,4.0)

Now, to find all possible links in $G_{new}$, we need to find all possible combination of {0, 1} in each space of $G_{new}$ matrix. But, for the large values of M, we need to take only random sample of this space to generate data because there will be large number of possible links which can't be viable to determine in real life for specific computing machine.

So, the number of network will be $(2C_1)^{(N+M)\times(N+M)}$ for all possible links in $G_{new}$ but most of them are not of $X_{in} = X_{in_{predicted}}$ and $X_{out} = X_{out_{predicted}}$

Let, assume these possible network are { $G_{possible}1, G_{possible}2,$ ................., $G_{possible}(2C_1)^{(N+M)\times(N+M)}$}

Now, to find possible network with $X_{in} = X_{in_{predicted}}$ and $X_{out} = X_{out_{predicted}}$, we need to test { $G_{possible}1, G_{possible}2,$ ................., $G_{possible}(2C_1)^{(N+M)\times(N+M)}$} in ANN2 to predict the Group.

If the any network from {$G_{possible}1, G_{possible}2,$ ................., $G_{possible}(2C_1)^{(N+M)\times(N+M)}$} get prediction accuracy for $Group_{valid\ network}$ > 0.80 then we can assume that this network will be network with $X_{in} = X_{in_{predicted}}$ and $X_{out} = X_{out_{predicted}}$ but these are the set of valid network with $G_{new}$ , $G_{new\ with\ hidden\ links\ and\ mising\ nodes}$ and other network with $X_{in} = X_{in_{predicted}}$ and $X_{out} = X_{out_{predicted}}$ .

Let, assume these are the network are $G_{valid} = \{G_{valid}1, G_{valid}2, .........\}$ where

$$G_{valid} \in \left\{ G_{new}, Or,\ G_{new\ with\ hidden\ links\ and\ mising\ nodes}\ Or,\ other\ networks\ with\ (X_{in} = X_{in_{predicted}}\ and\ X_{out} = \right.$$

So, we need to separate $G_{new\ with\ hidden\ links\ and\ mising\ nodes}$ from $G_{valid}$. For this, we need to find out those networks which all 1's position in $G_{new}$ exists in $G_{valid}$.

If each "1"s position in $G_{new}$ exists in the same position in $G_{valid}$ then we can say, that $G_{valid}$ is either $G_{new}$ or $G_{new\ with\ hidden\ links\ and\ mising\ nodes}$.

From $G_{new}$ and $G_{new\ with\ hidden\ links\ and\ mising\ nodes}$, we can easily eliminate $G_{new}$

After eliminating "the networks without each "1"s position in $G_{new}$ exists in the same position in $G_{valid}$" and $G_{new}$ from $G_{valid}$,

Only $G_{new\ with\ hidden\ links\ and\ mising\ nodes}$ will be remained in $G_{valid}$

So, the networks in $G_{new\ with\ hidden\ links\ and\ mising\ nodes}$ are the networks with hidden links and missing nodes for directed scale-free network.

## 3.3 The algorithm for training the neural networks

G ← directed scale free network on which we want to estimate the hidden links and missing nodes

$G_N$← Adjacent matrix of G

M ← number of missing nodes in $G_N$

$G_{new} \leftarrow Add(G_N, Unconnected(Node_{1,2,..,M}))$

## The size of $G_{new}$ is $(N + M)\times(N + M)$##

α ← {0.1,0.2,0.3,0.4,0.5}

β ← {0.1,0.2,0.3,0.4,0.5,0.6,0.7,0.8}

$\Upsilon \leftarrow \{0.1,0.2,0.3,0.4,0.5\}$

$X_{in} \leftarrow \{2.1, 2.2, 2.3, \ldots.., 2.9, 3.0\}$

$X_{out} \leftarrow \{2.1, 2.2, 2.3, \ldots.., 2.9, 3.0\}$

$\{G1, G2, G3, \ldots., G(100(N + M))\}$ ←directed Scale Free Graph Generation Algorithm (α, β, ϒ, $X_{in}, X_{out}$)

$\{Group1, Group2, Group3, \ldots., Group100\} \leftarrow Labeled(G1, \ldots.., G(100(N + M), label_by(X_{in}, X_{out}))$

$ANN1$ ←Artificial Neural Network (model=sequential)

$ANN1.\,input\ layer(number\ of\ node\ = (N + M) \times (N + M))$

$ANN1.\,hidden\ layer1(number\ of\ node\ = \frac{(N+M)\times(N+M)+100}{2},\ activation\ function\ = Rectifier\ Linear\ U$

$ANN1.\,hidden\ layer2(number\ of\ node\ = \frac{\frac{(N+M)\times(N+M)+100}{2}+100}{2},\ activation\ function\ = Rectifier\ Linear\ U$

$ANN1.\,output\ layer(\ number\ of\ node\ = 100,\ \ activation\ function\ = Softmax\ function)$

$ANN1.\,train\ (\ data\ = \{G1, G2, G3, \ldots., G(100(N + M)\},\ label\ = \ \{Group1, Group2, Group3, \ldots., Group10$

$Group_{predicted} \leftarrow ANN1.\,Prediction(maximum\ accuracy(Group1, Group2, Group3, \ldots., Group100))$

$X_{in_{predicted}} \leftarrow \ Group_{predicted}.\ X_{in}$

$X_{out_{predicted}} \leftarrow \ Group_{predicted}.\,X_{out}$

$ANN2$ ←Artificial Neural Network (model=sequential)

ANN2.$\,input\ layer(number\ of\ node\ = (N + M) \times (N + M))$

$ANN2.\,hidden\ layer1(number\ of\ node\ = \frac{(N+M)\times(N+M)+2}{2},\ \ activation\ function\ = Rectifier\ Linear\ Uni$

$ANN2.\,output\ layer(\ number\ of\ node\ = 2,\ \ activation\ function\ = Softmax\ function)$

$Group_{valid\ network} \leftarrow Labeled(G1, \ldots.., G(100(N + M), label(X_{in} = X_{in_{predicted}}\ ,\ \ X_{out} = X_{out_{predicted}}))$

$Group_{invalid\ network} \leftarrow Labeled(G1, \ldots.., G(100(N + M), label(X_{in} \neq X_{in_{predicted}}\ ,\ \ X_{out} \neq X_{out_{predicted}}))$

$ANN2.\,train\ (\ data\ = \{G1, G2, G3, \ldots., G(100(N + M))\},\ label\ = \ \{\ Group_{valid\ network},\ \ Group_{invalid\ network}$

$(\{G_{possible}1, G_{possible}2, \ldots\ldots\ldots, G_{possible}(2C_1)^{(N+M)\times(N+M)}\}$ $\leftarrow$ All possible combination of $\{0, 1\}$ in each space of $G_{new}$ matrix.

$$\{G_{valid}1, G_{valid}2, \ldots\ldots\} \leftarrow ANN2.Prediction(accuracy_for_Group_{valid\ network}(\{G_{possible}1, G_{possible}2, \ldots\ldots\ldots, G_{possible}(2C_1)^{(N+M)\times(N+M)}\} > 0.80)$$

$G_{new\ with\ hidden\ links\ and\ mising\ nodes} \leftarrow$ {The set of networks of $G_{valid}$ where each "1"s position in $G_{new}$ exists in the same position in $G_{valid}$ } $-$ $\{G_{new}\}$

Here, the networks in $G_{new\ with\ hidden\ links\ and\ mising\ nodes}$ are the networks with hidden links and missing nodes for directed scale free network.

For implementing the algorithm and performing the computational experiments, we used Python Module Networkx2.4 [19] for random graph generation algorithm and TensorFlow2.0 [20] for Artificial Neural Network.

## 4. Acknowledgement

This research was conducted with the help of Prof. Ljupčo Todorovski, University of Ljubljana, SI.

## 5. Conclusion

Though our algorithm is capable of successful estimation of missing links and hidden nodes in scale free networks, there are few limitation exists in this algorithm. First of all, we need to define how many missing nodes we want to find in given scale-free network. If any scale-free network has a large number of missing nodes, our algorithm may capable of finding those large number of missing nodes by trying to find sequentially from 1 missing node to ∞ missing nodes but this is not sufficient to find out those missing nodes due to some computational complexity, specially time and space complexity. Secondly, in real life, some of the scale free network sometimes don't follow the ideal properties of scale free network, in that case, our algorithm might not be able to find real life missing nodes and hidden links in scale free network. Thirdly, in this algorithm, we didn't analyze any worst case scenario for computational complexity. In case of any large number of nodes with large number of missing nodes in scale free network for a specific limited computing machine resources, worst case scenario can be happened. It is much appreciable in future research to estimate the computational complexity of this algorithm to find out the minimum computational machine resources to skip worst case scenario for large number of nodes in given scale-free network (described in section 3.2).

Besides few limitations, our algorithm is a good starting point and a proof of principle experiment for using neural networks to predict hidden links and missing nodes in scale-free networks. However, to prove its utility, we have to perform a series of computationally expensive experiments, which are beyond the scope of this seminar paper.